\documentclass[aps,prl,twocolumn,superscriptaddress,showpacs]{revtex4}
\usepackage{graphicx}

\begin{document}

\title{Fermi surface and quasiparticle dynamics of Na$_{0.7}$CoO$_2$ investigated by Angle-Resolved Photoemission Spectroscopy}
\author{M.Z. Hasan}\affiliation{Joseph Henry
Laboratories, Department of Physics, Princeton University,
Princeton, NJ 08544}\affiliation{Princeton Center for Complex
Materials, Princeton Materials Institute (PMI), Princeton, NJ
08544}\affiliation{Advanced Light Source, Lawrence Berkeley
National Laboratory, Berkeley, Ca 94720}

\author{Y.-D. Chuang}
\affiliation{Joseph Henry Laboratories, Department of Physics,
Princeton University, Princeton, NJ 08544}\affiliation{Advanced
Light Source, Lawrence Berkeley National Laboratory, Berkeley, Ca
94720}
\author{A. Kuprin}
\affiliation{Joseph Henry Laboratories, Department of Physics,
Princeton University, Princeton, NJ 08544}\affiliation{Advanced
Light Source, Lawrence Berkeley National Laboratory, Berkeley, Ca
94720}
\author{Y. Kong}
\affiliation{Joseph Henry Laboratories, Department of Physics,
Princeton University, Princeton, NJ 08544}
\author{D. Qian}
\author{Y.W. Li}
\affiliation{Joseph Henry Laboratories, Department of Physics,
Princeton University, Princeton, NJ 08544}
\author{B. Mesler}
\author{Z. Hussain}
\author{A.V. Fedorov}
\author{R. Kimmerling}
\author{E. Rotenberg}
\author{K. Rossnagel}
\author{H. Koh}
\affiliation{Advanced Light Source, Lawrence Berkeley National
Laboratory, Berkeley, Ca 94720}
\author{N.S. Rogado}
\author{M.L. Foo}
\author{R.J. Cava}
\affiliation{Princeton Center for Complex Materials, Princeton
Materials Institute (PMI), Princeton, NJ
08544}\affiliation{Department of Chemistry, Princeton University,
Princeton, NJ 08544}

\date{\today}


\begin{abstract}
We present an angle-resolved photoemission study of
Na$_{0.7}$CoO$_2$, the host material of the superconducting
Na$_x$CoO$_2\cdot y$H$_2$O series. Our results show a large
hexagonal-like hole-type Fermi surface, a strongly renormalized
quasiparticle band, a small Fermi velocity and a large Hubbard-U.
Along the $\Gamma\rightarrow$M high symmetry line, the
quasiparticle band crosses the Fermi level from M toward $\Gamma$
consistent with a negative sign of effective single-particle
hopping \textit{\textbf{t$_{eff}$}} (estimated to be about 8$\pm$2
meV) which is on the order of exchange coupling
\textit{\textbf{J}} in this system. Quasiparticles are well
defined only in the T-linear resistivity regime. Small single
particle hopping and unconventional quasiparticle dynamics may
have implications for understanding the unusual behavior of this
new class of strongly correlated system.
\end{abstract}

\maketitle

Since the discovery of cuprate superconductivity \cite{1}, the
search for other families of superconductors that might supplement
what is known about the superconducting mechanism of doped Mott
systems has been of great interest. The recent report of
superconductivity near 5K in the triangular lattice, layered
sodium cobalt oxyhydrate, Na$_{0.35}$CoO$_2\cdot$1.3H$_2$O,
suggests that such materials may indeed be found \cite{2}. The
crystal structure of this material, and its precursor hosts,
consists of electronically active triangular planes of edge
sharing CoO$_6$ octahedra separated by Na (and hydration) layers
that act as spacers, to yield electronic two-dimensionality, and
also as charge reservoirs\cite{2,3,4}. It has been argued that
Na$_x$CoO$_2$ is probably the only system other than cuprates
where a doped Mott insulator becomes a superconductor, although
the fully undoped system has not yet been realized \cite{5,6}.
Na$_{0.7}$CoO$_2$ is the host compound for these materials from
which Na$_x$ is varied to achieve superconductivity.

Despite some similarities with the cuprates, cobaltates show their
own unique set of anomalous properties. For Na$_{0.7}$CoO$_2$ the
anomalous Hall signal shows no saturation to 500 K \cite{7}, the
thermopower is anomalously high and magnetic field dependent
\cite{8}, there is linear-T resistivity (deviation from
Fermi-liquid behavior) from 2 K to about 100 K \cite{7} and there
is strong topological frustration\cite{5,6,9}. This class of
systems may potentially contain new fundamental many-body physics
and may as well be the realization of Anderson's original
triangular lattice RVB system \cite{10}. Therefore it is of
interest to have a detailed look at the charge dynamics, starting
with the host material. In this Letter, we report results of an
angle-resolved photoemission (ARPES) study of Na$_{0.7}$CoO$_2$. A
direct measurement of detailed Fermi surface topology and
quasiparticle dynamics is of significant interest as it would
provide a microscopic basis for understanding the complex electron
behavior. We observe a large hole-type hexagonal-like Fermi
surface. An extremely flat quasiparticle band is also found, which
hardly disperses more than 75 meV. The quasiparticle weight
decreases on raising the temperature to around 120 K and
disappears where the T-linear behavior in resistivity disappears.

Single crystals of Na$_{0.7}$CoO$_2$ were grown by the flux method
\cite{7}. Measurements were performed at the Advanced Light Source
Beamlines 12.0.1 and 7.0.2 using a Scienta analyzer. The data were
collected with 30 eV or 90 eV photons with better than 30 meV
energy resolution, and an angular resolution better than 1\% of
the Brillouin zone. The chamber pressure was better than
5$\times$10$^{-11}$ torr. Cleaving the samples in situ at 16 K
resulted in shiny flat surfaces, characterized by optical
(laser-reflection) and low-energy electron diffraction (LEED)
methods to be flat, clean and well ordered with the same symmetry
as the bulk. No signs of surface aging were seen for the duration
of the experiments. Instead of cleaving several crystals to map
the complete Fermi surface topology, we have worked on an image
mode with fully motorized manipulator at BL7.0.2 ARPES endstation
\cite{11}. The entire Fermi surface topology could be mapped out
within several hours after the cleavage.

\begin{figure}[ht]
\center
\includegraphics[width=6.5cm]{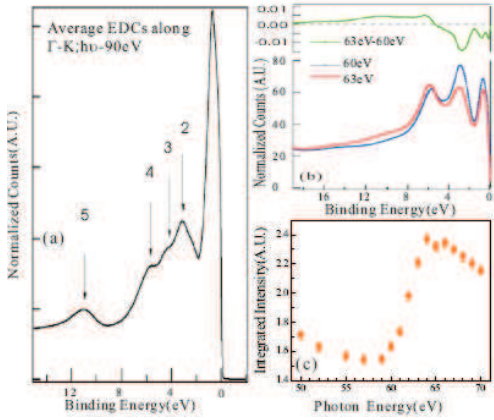}\caption{Valence Excitations and Resonance Behavior :(a)Full
valence band spectrum of Na$_{0.7}$CoO$_2$ taken at 90 eV photon
energy. It shows five prominent features - the feature at 0.7 eV
arises from Co t$_{2g}$ states. (b) Valence band near
Co:3p$\rightarrow$3d resonant excitation ($\sim$63 eV, red curve)
compared with off-resonant excitation ($\sim$60 eV, blue curve).
Resonance behavior of valence excitations shows the enhancement of
the 11 eV feature (green curve)- a correlation satellite. (c)
Incident energy dependence of the 11 eV feature. It shows a
Fano-type interference \cite{15}}
\end{figure}

Fig-1(a) shows the full valence band spectrum taken at 90 eV
photon energy. It shows five prominent features: 0.7 eV, 3 eV
(\#2), 4.1 eV (\#3), 6 eV (\#4) and 11 eV (\#5). The width of the
main valence density of states being on the order of 7 eV is
consistent with LDA calculations \cite{12}. To account for the
valence satellite one needs to consider the on-site Coulomb
interaction (U) and cluster calculations provide a measure of U.
Such calculations, with a strong Hubbard-U, suggest that the
valence band has predominantly $^1$A$_1$ character (low-spin
configuration t$_{2g}^x$e$_g^0$) and account for the 11 eV
feature. We identify the 11 eV feature as a correlation satellite
analogous to that seen in LaCoO$_3$\cite{13}, an indication of
strongly correlated behavior with Hubbard-U $\sim$ 5 eV. The major
low-energy feature is a broad band centered around 0.7 eV. Its
resonance and photon-energy dependent behavior suggest that it
consists of Co t$_{2g}$ derived states, consistent with LDA and
cluster calculations. Resonance behavior near $3p\rightarrow3d$
excitation is shown in Fig.-1(b). Valence band near Co:
$3p\rightarrow3d$ resonant excitation ($\sim$63 eV) is compared
with off-resonant excitation ($\sim$60 eV) by normalizing the area
under both spectra to unity. Resonance behavior of valence
excitations shows the enhancement of the broad 11 eV feature - a
correlation satellite. Fig-1(c) shows the detail resonance
behavior (Fano-type interference) of the correlation satellite. We
also observe enhancements of the valence satellites under
$2p\rightarrow3d$ resonance (not shown here), supporting the
identification of the 11 eV feature as a correlation satellite.
Existence of such features and the resonance behavior are strong
evidences for the system's highly correlated nature with large
Hubbard-U similar to the cuprates \cite{14,15}. Further details of
the resonance behavior would be reported elsewhere.

Much lower in energy near the Fermi level we observe a highly
momentum (\textbf{k}-)dependent quasiparticle feature. Fig-2(a)
shows this feature near the Fermi level crossing from M toward the
$\Gamma$ point in the Brillouin zone. Fig-2(b) shows part of the
energy dispersion curves (EDCs) corresponding to Fig-2(a). This
feature is well defined in momentum and energy and only weakly
dispersive, hardly dispersing more than 80 meV. Beyond 75 meV the
feature gets so broad (much broader than resolution) that the
quasiparticle is not defined anymore in the sense that its
lifetime is extremely short. To give a measure of bandwidth we
first make a dispersion plot using the following procedure : take
each EDC and subtract a background as shown in Fig.-2(c). Note
that a step-like background is observed for all k as evident from
Fig.-2(a). After background subtraction we take the centroid of
the peak for its energy value (Fig.-2(c)). Based on these peak
positions we make dispersion plots (Fig.-2(d) and (e)). To get a
measure for bandwidth we extrapolate the band to the zone boundary
as shown in Fig.-2(d) and (e). Since the band is narrow its
extrapolation to the zone boundary is within the error bars of the
measurement. This gives a value of about 70 $\pm$10 meV. Just
looking at the raw data (Fig.-2(a)) one could argue that the band
is narrower than 100 meV. It is interesting to note that this band
is not well defined over the full Brillouin zone - a case similar
to the cuprates and other correlated systems (strong correlation
can lead to significant lifetime shortening away from Fermi level
). This can also be due to strong scattering by collective modes
such as phonons \cite{23}. A bandwidth of about 400-500 meV for
cuprates is derived in a similar extrapolation basis \cite{23}.
Note that the cobaltate band is about at least a factor of 5
narrower than the cuprates.
\begin{figure} \center
\includegraphics[width=8.5cm]{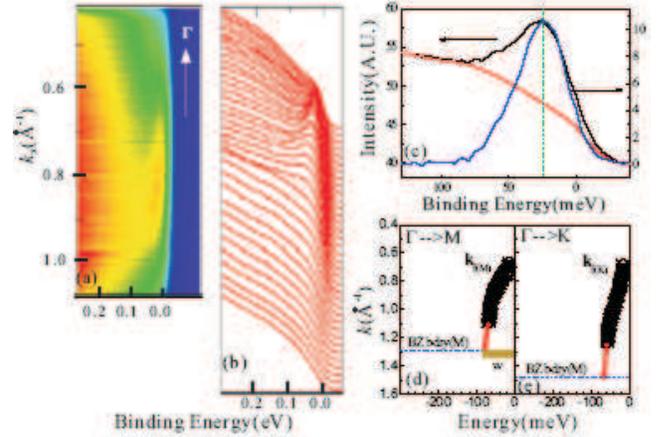}\caption{Quasiparticle dispersion : (a)$\Gamma\rightarrow$M
Fermi crossing. Color red reflects the highest intensity - yellow
to green to blue is in the order of decreasing intensity. (b) EDCs
corresponding to the image plot in (a). (c) A single EDC for k =
0.78 \AA$^{-1}$. To extract the peak position a background is
subtracted. A constant step-like background is best seen in Fig
2(a). Based on the extracted peak positions E vs. k plots are made
and shown in (d) for $\Gamma\rightarrow$M and in (e) for
$\Gamma\rightarrow$K directions.}
\end{figure}

Fig-3 shows the experimentally measured Fermi surface of
Na$_{0.7}$CoO$_2$ over the complete Brillouin zone measured at a
temperature of 16 K with a photon energy of 90 eV. The
frequency-integrated spectral distribution, the
n(\textbf{k})-image, has been taken by integrating spectral
weights within 75 meV below the Fermi level (and 25 meV above the
Fermi level). It reveals a large hole-pocket centered around the
$\Gamma$-point. The Fermi surface assumes a hexagonal character
with an average radius of 0.7 $\pm$0.05\AA$^{-1}$. This average is
calculated based on half of the Brillouin zone measured(Fig-3 is
symmetrized by two-fold). This is a bit larger than the size
calculated for the related compound NaCo$_2$O$_4$
(Na$_{0.5}$CoO$_2$) using LDA \cite{12}. If Na$_{0.0}$ corresponds
to half filling, Luttinger theorem suggests hole pocket area of
about 1/6 of the zone for the Na$_{0.7}$ case. The observed Fermi
surface area is a bit larger under this scenario. It may indicate
that a fraction of the doped electrons are in the Fermi sea, the
rest are presumably localized and form local moments as seen in
various experiments \cite{7,17,18}. The shape of the Fermi surface
in Na$_{0.7}$CoO$_2$ is rather hexagonal (anisotropic) in
character in agreement with LDA \cite{12}. We do not observe any
of the small satellite pockets predicted by LDA calculations
\cite{12} around this large hexagonal Fermi surface (at least for
the excitation photon energy of 90 eV where the search has been
most extensive). This could be due to strong correlation effects,
which can push the minority bands away from the Fermi level or
wash out their relative intensity. It could also be due to the
fact that one can not shift the chemical potential with doping in
a trivial way as commonly observed in doped Mott systems so a
straight forward rigid shift picture of the chemical potential for
comparison may not be appropriate.

Along the $\Gamma\rightarrow$M Fermi crossing the quasiparticle
band crosses the Fermi level from M toward $\Gamma$(Fig.2(a)) as
opposed to $\Gamma$ to M. Such a dispersion behavior is consistent
with the negative sign of single-particle hopping
(\textit{\textbf{t}}). Based on several EDC cuts we estimate the
total bandwidth of this system which is about 70$\pm$10 meV (less
than 100 meV). For a tight-binding hexagonal lattice total
bandwidth \textit{\textbf{W}} = 9\textit{\textbf{t}} where
\textit{\textbf{t}} is the nearest neighbor single-particle
hopping. This gives a value of the effective single-particle
hopping, \textit{\textbf{t$_{eff}$}} of about 8$\pm$2 meV in this
system \cite{19}. It is interesting to note that this value is on
the order of exchange coupling \textit{\textbf{J}} ($\sim$ 10
meV)\cite{17,18} in this system. This bandwidth is an order of
magnitude renormalization compared to mean field calculations,
which suggest a bandwidth of order 1 eV to 1.4 eV\cite{5,6} or
0.48 eV\cite{12}. Such enhancement of effective mass is in
agreement with electronic specific heat measurements \cite{20,21}.
Bandwidth suppression may also be responsible for enhancing the
thermopower by an order of magnitude\cite{8}. Single-particle
hopping being on the order of exchange coupling suggests that the
charge motion would be significantly affected by the spin dynamics
of this system. Also a small value of \textit{\textbf{t}} suggest
that this system has an unusually small fermion degeneracy
temperature compared to other metals.

Momentum distribution curves (MDCs) could be fitted with single
Lorentzian function, on top of a linear background. The MDC
fitting is known to produce reasonable results, especially when
features are close to Fermi level E$_f$\cite{22,23} though not
without controversies. Lorentzian lineshapes indicate that the
quasiparticle self-energy is weakly dependent on the momentum
normal to the Fermi surface \cite{23}. We can then estimate the
quasiparticle velocity normal to the Fermi surface. Such
\textit{MDC based} Fermi velocity is found to be less than 0.4
eV$\cdot$\AA. This value is much smaller than the nodal Fermi
velocity of the cuprates ( $\sim$ 1.5 eV$\cdot$\AA) extracted in a
similar way. This is consistent with the fact that the cuprate
bands are more dispersive \cite{23}.
\begin{figure}[ht]
\includegraphics[width=4.5cm]{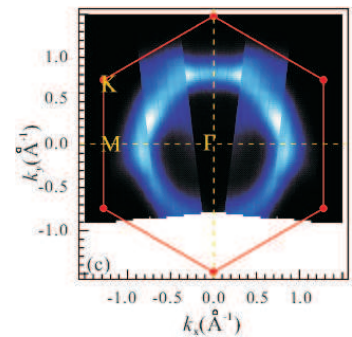} \caption{Fermi surface : n(k) plot generated by
integrating within 75 meV of Fermi level. A large hole-pocket is
centered around the $\Gamma$-point. The Fermi surface, exhibiting
some hexagonal anisotropy, is the inner edge of pocket as shown
over the complete Brillouin zone.}
\end{figure}
We further studied the quasiparticle spectral weight near the
Fermi level as a function of temperature (Fig.-4). The
quasiparticle weight decreases to almost zero (background level)
on raising the temperature to around 120 K (Fig.4(b)). This
roughly coincides with the temperature where the T-linear behavior
in resistivity gives way to stronger T-dependence (Fig.4(b)).
Apparently, quasiparticles exist in this system only in the
temperature regime where resistivity is linear in T and transport
behavior is non-Fermiliquid like. This is in contrast with the
conventional expectation that well defined quasiparticles are
signatures of good Fermi liquid behavior. We also note that this
temperature scale coincides with our scale of effective
single-particle hopping \textit{\textbf{t}} $\sim$ 8 meV $\sim$
100 K. A dip is observed around 100 K in the temperature
dependence of Hall coefficient\cite{7}. This is also the
temperature range when thermopower starts to deviate from
linear-like behavior \cite{7}. We argue that this is an important
(and most relevant) energy scale to describe the physics of these
systems. A strong lack of quasiparticle weight at higher
temperatures is also reminiscent of the lack of quasiparticles in
the high temperature normal state of cuprates \cite{23}. But in
cuprates, it is difficult to study the normal state over a large
temperature range as it enters the superconducting state at a
fairly high temperatures. Also cuprates can not be too heavily
over doped. Cobaltates may offer the unique opportunity to study
the normal state behavior of a heavily doped Mott system. This
class of cobaltates has it's own uniqueness and hence interesting
in its own rights. For the quasiparticle behavior, it may be that
the transport in Na$_{0.7}$CoO$_2$ becomes incoherent well before
a dimensional cross-over(two to three dimensional charge
transport, \cite{17}) is reached. Such behavior could be related
to the highly frustrated nature of the antiferromagnetic
interactions in a triangular lattice.

Previously studied layered cobaltates structurally similar to the
cuprate family of Bi$_2$Sr$_2$CaCu$_2$O$_y$ such as
Bi$_2$Sr$_{2.1}$Co$_2$O$_y$ and
Bi$_{1.5}$Pb$_{0.5}$Sr$_{2.1}$Co$_2$O$_y$ were not reported to
exhibit a quasiparticle state \cite{24}. Cobaltates such as
(Bi$_{0.5}$Pb$_{0.5}$)$_2$Ba$_3$Co$_2$O$_y$ or NaCo$_2$O$_4$
exhibit quasiparticles and Fermi surfaces that were found to be
roughly rounded, but an anisotropic shape was not reported
\cite{16}. In case of  Na$_{0.7}$CoO$_2$  we find a hexagonal
character and an even larger Fermi surface. The Fermi velocity in
Na$_{0.7}$CoO$_2$ is found to exhibit anisotropy as much as 20\%.
A thorough comparison of Na$_{0.7}$CoO$_2$ with Na$_{0.5}$CoO$_2$
is difficult since detailed characterization of the latter has not
been reported \cite{16}. Unlike Na$_{0.5}$CoO$_2$, no dimensional
crossover (two to three dimensional charge transport) exists in
the Na$_{0.7}$CoO$_2$ up over to 300 K. The temperature dependence
of quasiparticles look somewhat similar, however, hence we rather
ascribe such temperature dependence in Na$_{0.7}$CoO$_2$ with the
T-linear resistivity (non-Fermiliquid) regime. There are also
dissimilarities at higher energies: the so called "broad hump"
\cite{16} around 0.7 eV is narrower and sharper (more than 500
meV) in Na$_{0.7}$CoO$_2$ than in Na$_{0.5}$CoO$_2$ for
momentum(k) values near Fermi surface as seen in Fig.-4(b).
Furthermore we see a strong temperature dependence of the 0.7 eV
feature in Na$_{0.7}$CoO$_2$ - it tends to move toward higher
binding energies at temperatures where the departure from a
T-linear resistivity grows. The overall temperature induced
spectral weight changes are found to be conserved within a range
of 1 eV (changes in the valence band near the Fermi level due to
changes in T occur only within 1 eV of the Fermi level). Such high
sensitivity of temperature over a higher energy band (in addition
to the quasiparticles) and large spectral weight redistribution is
a further signature of this system's strongly correlated behavior
\cite{25} in addition to its highly flat band character which is
typically the case for strong electron-electron correlation. A low
temperature phase transition has been reported for
Na$_{0.75}$CoO$_2$ -an indication that higher (commensurate)
doping drives the system towards an ordering instability
\cite{26}. Despite the possibilities of kinematic nesting
associated with the observed Fermi surface, the ordering in
Na$_{0.75}$CoO$_2$  may be more complex due to the strongly
correlated nature and the frustrated interactions in the system.
But it would be interesting to study the nature of short-range
dynamical correlations around q $\sim$ 3$\pi$/2 using inelastic
x-ray and neutron scattering techniques \cite{27}.
\begin{figure}[ht]
\includegraphics[width=8.5cm]{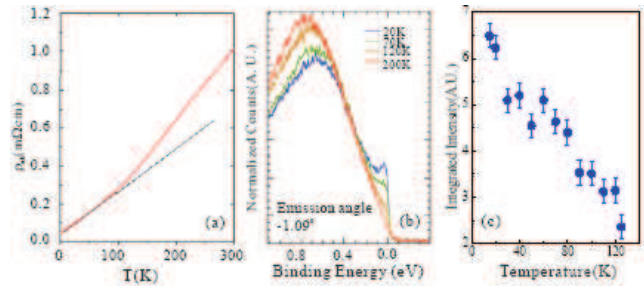} \caption{Temperature
Dependence : (a) In-plane resistivity in Na$_{0.7}$CoO$_2$ is
linear up to 100 K \cite{7} and then gradually crosses over to a
stronger T-dependence. (b) Temperature dependence of
quasiparticles near the  $\Gamma\rightarrow$M  Fermi crossing. The
quasiparticle spectral weight ceases to exist above 120 K, close
to the temperature where the T-linear behavior of the resistivity
disappears. (c) Background subtracted integrated quasiparticle
spectral weight is plotted as a function of temperature}
\end{figure}

In conclusion, we report an angle-resolved photoemission study of
Na$_{0.7}$CoO$_2$. Resonant scattering of valence excitations
indicate the existence of a large Hubbard-U supporting the
strongly correlated nature of the system. The low-energy results
reveal a large hexagonal-like hole-type Fermi surface and an
extremely narrow quasiparticle band - an order of magnitude
renormalization from the meanfield value. Such bandwidth
suppression may be the key to understand the enhancements of
thermopower and electronic specific heat. Effective single
particle hopping being on the order of exchange coupling suggests
that charge motion is significantly influenced by spin
fluctuations in these systems. Effective (small) hopping being on
the order of exchange coupling strongly indicates possible
existence of unconventional physics (including RVB phases).
Quasiparticles are well defined only in the T-linear resistivity
(non-Fermi liquid) regime. From a theoretical perspective, it
would be interesting to understand the emergence of this small
degenerate energy scale in the cobaltates. The system's strongly
correlated nature as manifested from this extremely narrow band
and large spin-entropy observed by transport measurements \cite{7}
taken together may shed clues to understand the broad spectrum of
unusual properties including superconductivity at low doping. It
would be interesting to study the doping evolution of the Fermi
surface and quasiparticle behavior of these systems by combining
ARPES and newly developed momentum resolved inelastic x-ray
scattering \cite{27}. Furthermore, any comprehensive theory for
cobaltates need to account for the low degenerate energy scale and
unconventional quasiparticle dynamics observed.

The authors thank P.W. Anderson, N.P. Ong, P.A. Lee, D.A. Huse,
P.M. Chaikin, A. Millis, S. Sondhi, J. Lynn, S. Maekawa, A.
Vishwanath and Y. Wang for valuable discussions. The experiment
was performed at the ALS of LBNL, which is operated by the U.S.
DOE's BES with contract DE-AC03- 76SF00098. MZH acknowledges
partial support through NSF-MRSEC (DMR-0213706) grant and R.H.
Dicke Award by Princeton Univ. Materials synthesis supported by
DMR-0213706 and the DOE, grant DE-FG02-98-ER45706. Email
correspondence : mzhasan@Princeton.edu

\end{document}